\documentclass[12pt]{iopart}
\begin{document}

\title[Very high frequency...]
  {Very high frequency gravitational wave background in the universe}
\author{G.S. Bisnovatyi-Kogan\dag\ and V.N.Rudenko\ddag}
\address{\dag\ Space Research Institute, Moscow, Russia, Profsoyuznaya 84/32,
  Moscow 117810, Russia and Joint Institute of Nuclear Researches, Dubna,
  Russia}
\address{\ddag\ Sternberg Astronomical Institute,MSU, Universitetski pr. 13,
Moscow, 119899 Russia}

\ead{gkogan@mx.iki.rssi.ru}


\begin{abstract}

Astrophysical sources of high frequency gravitational radiation
are considered in association with a new interest to
very sensitive HFGW receivers required for the laboratory GW
Hertz experiment. A special attention is paid to the phenomenon
of primordial black holes evaporation. They act like black body to
all kinds of radiation, including gravitons, and, therefore, emit
an equilibrium spectrum of gravitons during its evaporation.
Limit on the density of high frequency gravitons in the Universe
is obtained, and possibilities of their detection are briefly discussed.

\end{abstract}

\section{Introduction}

Search for gravitational waves radiated by extraterrestrial
sources is carried out now by  large experimental teams of
different countries. Cryogenic GW bars detectors are in a
permanent duty cycle of observations (Allen et al., 2000; Astone
et al., 2001), and first test series of observation were performed
recently on large gravitational interferometers (Abbott et al.,
2003; Shoemaker, 2003). A lot of experimental and theoretical
studies have been done in the process of development of this
branch of physics conventionally called ``Gravitational Wave
Experiment''. Detailed calculations had been done  of continuous
and pulse GW-signals emitted by different relativistic objects in
the Universe, and algorithms had been developed by many authors
(see for example the recent paper of
Cutler and Thorne, 2002) of optimal data
processing for filtering these signals from the detector's noise
background. Modern programs for GW search with both
ground based and space apparatus deal with frequency range which
does not exceed a few kilocycles per second. The upper frequency
limit for relativistic sources is determined by the inverse
time of fly-by a strong field
region $ \nu \leq (c/r_{g}) \sim 30 (3M_{\odot}/M) kHz $.
The lower frequency, the more intensive GW bursts one could expect
according to the modern astrophysical forecast; by in other words
massive relativistic stars $ M>3M_{\odot}$ are ``too heavy `` and
``too inertial'' for the production a powerful high frequency
radiation. Nevertheless  a new attention was recently attracted
just to the region of very high frequency gravitational waves
$10^{10}-10^{15}\, Hz $ in connection with the old problem of the
possibility of a laboratory GW Hertz experiment (HFGW ,2003; Rudenko,
2003). A stimulating idea is based on the understanding that in
the lab one could organize a coherent radiation from a large number of
elementary quadruples (atoms or molecules of properly selected
media, vortexes in a superconductors etc.). The radiated power is expected
to be large enough because the very small factor of the
quadruple formula $ G/c^{5}$ can be compensated by other factors:
the enormous number of participated quadruples up to
$10^{22}-10^{24}$ per $cm^{3}$,  sixth power of the oscillating
frequency $\omega^{6}$, and a very sharp beaming of the output
signal (Rudenko, 2003). In turn an elementary quadruple as a GW
detector at very high frequencies should be more effective because
its size can be matched to the gravitational wave length (at low
frequencies bar detectors and even interferometers have the
loose factor (dismatch) $l/\lambda_g\ll 1$).

Preliminary calculations (Pisarev, 1979; Akihhin, et al. 1985) had shown
that at least a couple ``GW generator-receiver'' might be thought more or less
realistic in the lab conditions at ``optical frequences'' $ \nu \sim 10^{14}-
10^{15} Hz $ where the generated GW power achieves $\sim 1$ erg/sec.
 From the applied and engineering point of view the
practical realization of the GW Hertz experiment would open a way to GW
communication systems and other applications
(HFGW, 2003).
In this paper, however, we follow the traditional approach in which the cosmic
gravitational radiation is considered as a new channel of astrophysical
information. So our goal here is to select conceivable sources of HFGW in
astrophysics and to discuss the possibility of their2 detection.

\section{The nature of HFGW in astrophysics}

There are not too many papers in scientific literature with a detailed analysis
of astrophysical objects and processes resulted in gravitational
radiation at very high frequencies including ``optical'' and ``gamma ray''
frequency ranges. A list of HFGW sources theoretically studied
more or less up to now is restricted by four different classes of objects:
a thermal gravitational radiation of stars,
mutual conversion of electromagnetic and gravitational waves in a magnetized
interstellar plasma, relic cosmological gravitational wave background and
gravitational radiation from very low mass primordial black holes. Below
we consider the first three sources very briefly, and the last one
in more details.

i)Thermal gravitational radiation of stars. \\
 Thermal motion of charged particles in the stellar high temperature
proton-electron plasma produces a gravitation radiation noise.
Estimation by the order of magnitude had been performed, for example,
in papers of Braginskii and Rudenko (1963), Weinberg (1972),
Galtsov and Gratz (1974).
One has to calculate more or less rigorously the output of gravitational
radiation associated with the elementary process of scattering of two
particles (approaching and flying away), and then to make summation over
all events at given concentration and velocity distribution of particles.

The most probable frequency of radiation  is determined by the
frequency of collisions $\nu_{c}\sim e^{4}n_{e} (kT)^{-3/2}m_{e}^{-1/2}
\sim 10^{15}$ Hz, and the highest frequency corresponds to the thermal
limit $ \omega_{m}=kT/\hbar \sim 10^{18}$ Hz (we use here the
plasma parameters in the center of the Sun:
the electron concentration $n_{e}=3\cdot 10^{25}$ cm$^{-
3}$, temperature $T=10^{7}$ K, here $ e, \, m_{e}
$ are the electron charge and mass respectively, and $\hbar $ is the Planck
constant).
The crude estimate of total ``thermal GW power'' can be written as
(Weinberg 1972)
\begin{equation}
P_{g}\sim (64G/c^{5})\,
n_{e}^{2}\,e^{4}(kT/m_{e})^{1/2}\,\omega_{m}V_{\odot}\sim
10^{14}-10^{15} {\rm erg/s}.
\label{1}
\end{equation}
For the numerical estimate we use the volume of stellar core
$V_{\odot}= 10^{31}-10^{32}\, {\rm cm}^3$. The result (\ref{1}) forecasts
that at the Earth the ``thermal gravitational noise of the Sun'' provides
a stochastic flux at ``optical frequencies'' of the order $\sim (10^{-13}-
10^{-14})$ erg/sm$^2$/s. \\
In the paper of Galtsov (1975) \cite{gal} it was calculated that
at more high frequencies $\nu \geq 10^{17}$ Hz (the x-ray range)
the expected GW output might be larger at two orders of value due
to the "photocoulomb processes", i.e. the scattering of photons at
charged particles with the birth of
gravitons. \\
It is interesting to note that
such a level of GW noise is comparable with the GW energy flux from binary
stars of the
Galaxy at very low frequencies: so the WU Ma (period 0.33 d, distance
110 pc) provides $\sim 3\cdot 10^{-13}$erg/cm$^{2}$/s and WZ Sge (period
81 min , distance
100 pc) gives $\sim 4\cdot 10^{-13}$ erg/cm$^{2}$/s. According to calculation
of Mironovskii (1965) the integral radiation of all Galaxy binaries have a
stochastic character with a peak at period $\sim 4$ hours, corresponding to
the flux $ 10^{-7}-10^{-8}$ erg/cm$^{2}$/s. Similar
accumulation has to take place for the ``thermal gravitational noise of stars''
but due to large distances to stars the integral radiation
cannot exceed the Solar effect. By the order of magnitude the integral
gravitational flux from galactic stars
at ``optical region'' might achieve only $10^{-16}$ erg/cm$^{2}$/s (for
an average
distance of 100 pc, and N=$10^{9}$),  although possible effects of close
local anomalies in star space distribution has to be addressed.

ii) Graser beams in interstellar plasma.\\
Possibility of coherent generation of HFGW by electromagnetic waves
propagating through an interstellar plasma and magnetic fields initially was
considered
in the pioneering papers of Gertsenshtein (1962) and Zeldovich (1973). After
that many theoreticians had analyzed details of this process in lab and
astrophysics. As a last development one could look at the paper of Servin
and Brodin (2003), and reference therein, complete theoretical calculations
may be found in the book of Galtsov et al. (1984).

Transformations between gravitational and electromagnetic waves is impossible
in empty (flat) space but takes place in the presence of ionized media,
magnetic fields (or curvature). It creates condition for a parametric
interaction these waves with an important sequence of effective energy
exchange.

Under special matching conditions ("wave synchronization" or "equality
of phase velocities") EM waves generate the GW beam which is amplified
by the coherent quadruple oscillations of atoms (as well as of electrons or
molecules) of the media driven by the EM wave. If a coherent length
is long enough the radiated and sharply beamed GW power might achieve
a remarkable level. In particular, a propagation of the powerful EM wave
in a plasma cylindrical "waveguide"  $(r,l;\,l\gg r)$  will produce a
double frequency GW wave at the Cherenkov angle $\Theta\sim\arccos(v_p/c)$
(where $v_p$-is the phase velocity of the EM wave along the "waveguide").
The upper limit of the generated GW power was calculated by Galtsov et. al.
(1984).
For a relativistic plasma with the
characteristic parameter $\gamma=(eE/mc\omega)\sim 1$ which might be
produced by very intensive EM optical flux $F\sim 10^{19}
{\rm W/cm}^2 {\rm sec}$ (the corresponding field amplitude
$E\sim 10^{10} {\rm V/cm}$)
the output GW power per unit plasma volume is estimated as
\begin{equation}
 P_{g}\sim\frac{G}{c}\gamma^2 (m_{e}\omega n_{e})^{2}(r^{2}l)(\lambda/r)^{2}
\sim 10^{-5} {\rm erg/cm}^3 s,
\label{2}
\end{equation}
the following parameters were used above for the numerical estimate: \\
$ \omega=10^{15}$ rad/sec, $\lambda=10^{-4}\,$cm, $n_{e}=10^{22}\,$cm$^{-3},\,
\, r \sim l\sim 1\,$cm.

In astrophysics such condition might exist in the envelope of supernova or
hypernova (gamma-ray sources), jets from accreting dense objects, in the
vicinity of neutron stars etc. In fact this type of GW generators at "optical
frequencies" is similar to a well
known phenomena of EM "cosmic masers" at GHz and THz region
(Bochkarev 1992), and may be called as "gravitational masers" or
"grasers". If the length of the "coherent plasma waveguide" is
comparable with the scale of cells of radio masers $\sim 10^{16}-
10^{17}$ cm (Bochkarev 1992), the power of the generated GW beam
could achieve $10^{11}-10^{12}$ erg/s, together with a sharp beaming.
A search for such GW long living or transient sources  might be carried out
with detectors operating in a manner of "reverse scheme" when in a
nonlinear "opto-acoustical medium" an incoming HFGW produces a weak
electromagnetic wave responce (Akishin et al.,1985; Galtsov et al., 1984).

iii)Relic gravitational wave background \\
In the frame of the standard cosmological model, the relic stochastic
gravitational wave background must exist at present as a result of
amplification of initial GW zero quantum fluctuation by variable
gravitational field of the expanding Universe (Grishchuk, 1988, 2001, 2003).
Such a background exists also in the inflationary and cosmic string cosmologies
with some differences concerning the forecasted intensity and spectral
features (Allen, 1997).
All theories give a non-thermal energy density spectrum for the GW
relic background with a growth at low frequencies and a fall in the high
frequency region.
It reflects a specific parametric mechanism of the GW amplification.
Gravitons with wavelengths of the order of the scale factor get a more
effective pump from the gravity field of the expanding world.

The spectrum at high frequencies has a cutoff approximately in the region
$\nu_{c}\sim 10^{11}$ Hz; it corresponds to the temperature $0.9$ K which
GW relic background
would have in the case of the adiabatic evolution (Allen, 1997). Beyond
$\nu_{c}$ the spectrum falls down very quickly (some cosmic strings
theories give the cutoff shifted to $10^{13}-10^{14}$ Hz). So the
"optical" or "gamma-ray" gravitons practically has to be absent in the
relict background. In its "radio frequency" part the parametrical
GW energy density spectrum  follows approximately the law $ \Omega_{gw}\sim
h^{2}(\nu/\nu_{H})^{2}$, where as usual $h$ is the metric perturbation,
$\Omega_{gw}=\rho/\rho_{c}$ is the normalized GW energy density,
$\nu_{H}\simeq 10^{-18}$ is the Hubble frequency. At the cutoff frequency
$10^{11}Hz$ the intensity estimation is $h\sim 10^{-32}$ for
the metric perturbation and $F_{gw} \sim 10^{-5}-10^{-6}$ erg/cm$^{2}$/s
for the GW flux. These data are very small in comparison with the sensitivity
typical
of modern gravitational detectors $h\sim 10^{-24}-10^{-25}$ at the frequency
range $(10^{2}-10^{3})$ Hz. In this region there is the only hope
of registering the relict background with an advanced version of GW
interferometers (Grishchuk et al. 2001, Allen, 1997).\\

iiii) GW radiation from primordial black holes.\\
The high frequency GW $\nu \geq 10^{10}$ Hz could be radiated by
relativistic objects with a small mass and gravitational radius
$r_{g}\leq 3$ cm, i.e. it
might be black holes
with $M\leq 10^{-5}M_{\odot}$. It was supposed such black holes
might exist as a
primordial dark matter created at the very early Universe
(Carr, 1976; Zeldovich 1980).
In the frame of this hypothesis one could speculate over an
evolutionary scenario of these objects associated with GW radiation. It
could be done for example in the manner of the paper of Nakamura et
al.(1997), where the authors supposed an existence binaries composed by MACHO
objects,
primordial black holes with $M \sim 0.5M_{\odot}$, and then calculated
GW radiation from the coalescence of such binaries. As a part of dark
matter these objects may be located in the close environment of the Galaxy
increasing the number of
expected coalescences per year. The estimations of Nakamura et al. (1997)
gave the probability $5\,10^{-2}$ events/year/gal, which is at least two
order of magnitude larger than for conventional relativistic binaries
(Cutler and Thorne 2002). A hypothesis of binaries composed by mini
black holes with much smaller masses (so called ``holeums'') was considered
recently by Chavda and  Chavda, 2002).

However there is an obvious mechanism for HFGW production based on
the effect of evaporation of black holes of small masses through the
"graviton degree of freedom".
In the rest part of this paper we will try to estimate an efficiency of
this channel which
might compensate the deficit of high frequency gravitons in the relic
gravitational wave background. At first we remind briefly the main
features of the effect. \\

It was shown by Hawking (1975), following a thermodynamic analysis
of Bekenstein (1973), that black hole has properties of a black
body with a temperature $T_{\rm H}$.

\begin{equation}
\label{eq1}
 T_{\rm H}=\frac{\hbar c^3}{8\pi kGM_{\rm bh}}=6.2
 \cdot 10^{-8}\left(\frac{M_{\odot}}{M_{\rm bh}} \right) K.
\end{equation}
The temperature $T_{\rm H}$ may be estimated phenomenologically from the
obvious assumption, that the characteristic length of the emitting
radiation is of the order of the gravitational radius of the black
hole. Assume that the wave is not absorbed when one quarter of its
length exceeds the gravitational radius, so that
$\lambda_{lim}=\frac{8GM_{\rm bh}}{c^2}$. The corresponding frequency and
quantum of energy are

$$\nu_{lim}=\frac{c}{\lambda_{lim}}=\frac{c^3}{8GM_{\rm bh}},
\quad \varepsilon_{lim}=2\pi \hbar \nu = \frac{\pi}{4}
\frac{\hbar c^3}{GM_{\rm bh}}.$$
The black body may not absorb quanta with
wavelength $\lambda > \lambda_{lim}$, so it may emit them,
see also Kuchiev (2003).
The black body emits the planckian spectrum of radiation, whose
maximum energy $\varepsilon_m$ is connected with the temperature
$T$ as $\varepsilon_m=2.822 kT$ (Landau and Lifshitz, 1995). While
almost all quanta with $\varepsilon < \varepsilon_{lim}$ are
emitted by the black body, the limiting energy should be much
greater than the energy of the maximum $\varepsilon_m$,
$\varepsilon_{lim}$=$\eta \varepsilon_m$,
$T=\frac{\varepsilon_{lim}}{2.822 \eta k}$, with $\eta \gg 1$.
Taking $\eta=\frac{2\pi^2}{2.822}=7.0$, we come to the
Hawking temperature (\ref{eq1}).

\section{Energy spectra of emitting particles}

The black body spectrum of the emitting gravitons is
described by the Planck formula (Landau and Lifshitz, 1995). Like
photons, the massless gravitons have two independent polarizations, and
therefore have the same statistical weight $g_{\rm gr}=g_{\rm ph}=2$,
in spite of twice larger spin $s_{\rm gr}=2$, $s_{\rm ph}=1$.
The radiation of the black hole at low frequencies $h\nu \ll kT_{\rm H}$
is overestimated by the Planck formula, but we will ignore this.

\begin{equation}
\label{eq2}
\frac{dE_{{\rm gr},\nu}}{d\nu}=
\frac{g_{\rm gr}h^4}{2\pi^2\hbar^3 c^3}\frac{\nu^3}{e^{h\nu/kT}-1}\,\,
{\rm erg/cm^3/Hz}.
\end{equation}
The spectral density $F_{{\rm gr},\nu}$ of the flux of gravitons from
the unit surface of the black hole is

\begin{equation}
\label{eq3}
F_{{\rm gr},\nu}=\frac{c}{4} \frac{dE_{{\rm gr},\nu}}{d\nu}\,\,
{\rm erg/sm^2/s/Hz}.
\end{equation}
The total equilibrium energy  density $E_{{\rm gr}}$ and flux $F_{{\rm gr}}$
of gravitons is obtained after the integration over the frequency

\begin{equation}
\label{eq4}
E_{{\rm gr}}=\frac{2 g_{\rm gr}}{c}\sigma T^4, \quad
F_{{\rm gr}}=\frac{g_{\rm gr}}{2}\sigma T^4, \quad
\sigma=\frac{\pi^2 k^4}{60 \hbar^3 c^2}.
\end{equation}
Decrease of the mass of a black hole due to its evaporation and emission
of massless or ultrarelativistic particles is written as

\begin{equation}
\label{eq5}
\frac{dM_{\rm bh}}{dt}=-(\Sigma_{\rm boz}g_i+\frac{7}{8}\Sigma_{\rm fer}g_i)
\frac{\sigma T^4_{\rm H}}{2c^2} 4\pi\left(\frac{2GM_{\rm bh}}{c^2}\right)^2,
\end{equation}
where the sums ($\Sigma$) are taken over the statistical weights of all kinds
of the particles, bozons and fermions. For the sum of photons,
gravitons, 3 types of massless neutrino (antineutrino) and their massive
leptons, we have, using (\ref{eq1}), and the equality
$g_{\rm ph}+g_{\rm gr}+3\,\frac{7}{8}\,(g_{\nu}+g_{\tilde \nu}+g_{\rm e^-}
+g_{e^+})=2+2+3\,\frac{7}{8}\,(1+1+2+2)=\frac{79}{4}$,

\begin{equation}
\label{eq6}
\frac{dM_{\rm bh}}{dt}=-158\pi\frac{G^2M^2_{\rm bh}}{c^6}\sigma T^4_{\rm H}=
-\frac{79}{8^4 30\pi}\frac{\hbar c^4}{G^2 M^2_{\rm bh}}.
\end{equation}
Solving (\ref{eq6}) we find a time $t_{\rm bh}$ of the black hole
evaporation. The main portion of the energy, and main evaporation time
is collected at initial stages.
We take (\ref{eq6}) for all masses, while at $kT_{\rm H} >\sim
0.3 m_{\pi}c^2\approx 100$ MeV pions started to be radiated, and at
larger $T_{\rm H}$ other massive particles,
up to quarks and gluons are emitted (Barrau et al., 2003).

\begin{equation}
\label{eq7}
t_{\rm bh}=\frac{8^4 10 \pi}{79} \frac{G^2 M_0^3}{\hbar c^4}=
8.50\cdot10^{18}M_{15}^3,\quad M_{15}=\frac{M_{\rm bh}}{10^{15} {\rm g}}.
\end{equation}
Black holes evaporate during the cosmological time $5 \cdot 10^{17}$ s when
its initial mass $M_0$ is less than $M_b=4\cdot 10^{14}$ g.
(Page, 1976; Novikov and Frolov, 1998). Its temperature then exceeds

\begin{equation}
\label{eq8}
T_{\rm H}>\, 3.1\cdot 10^{11}\,\,{\rm K}\,\,=\, 26.6 \,{\rm MeV}.
\end{equation}
We see than that we may neglect pions and barion-antibarion emission at the
initial stages of evaporation of black holes with $M \ge M_b$. We also have
taken ultrarelativistic
Fermi distributions for all leptons, overestimating the radiation of $\tau$-
leptons.
While we neglect losses, connected with other particles, like pions, we may
expect that this approximation gives reasonable result for the black
hole evaporation rate.

Let us find the spectrum of the gravitons or photons (bosons)
${\cal E}(\nu)$, produced at the
evaporation of the black hole with the initial mass $M_0$.
We integrate (\ref{eq3}) over the whole evaporation time, which is
reduced to the integration over the mass, using (\ref{eq6}). We have

\begin{equation}
\label{eq9}
{\cal E}(\nu)=\frac{15}{79 \pi^5} \frac{\hbar c^5}{G} \frac {h}{(h\nu)^2}
\int_0^{x_0} \frac{x^4 dt}{e^x-1}\,\, {\rm erg/Hz}\,,
\end{equation}
$$x=\frac{8\pi GM_{\rm bh} h\nu}{\hbar c^3},\,\,\,
x_0=\frac{8\pi GM_0\, h\nu}{\hbar c^3}.$$
The spectrum (\ref{eq9}) has a maximum at

\begin{equation}
\label{eq10}
h\nu_m=\frac{\hbar c^3\, x_{m{\rm b}}}{8 \pi GM_0},
\end{equation}
where $x_{m{\rm b}}$ is a solution of the integral equation

\begin{equation}
\label{eq11}
\frac{x_{m{\rm b}}^5}{e^{x_{m{\rm b}}}-1}=2\int_0^{x_{m{\rm b}}}\frac{x^4}
{e^{x}-1}dx,\,\,\,x_{m{\rm b}}=3.933350238.
\end{equation}
In the approximation that the black hole emits the planckian spectrum,
the maximum of the radiation spectra of photons and gravitons corresponds
to the energy $h\nu_m$ and is connected with the Hawking
temperature of the initial mass black hole $T_{{\rm H}0}$ as

\begin{equation}
\label{eq12}
h\nu_m=x_{m{\rm b}}\,kT_{{\rm H}0}=3.933\,kT_{{\rm H}0},
\end{equation}
which is not far from the maximum of the Planck spectrum (Landau and Lifshitz,
1995) $\nu_{{\rm P}m}$, corresponding
to one temperature $T$: $h\nu_{{\rm P}m}=x_{{\rm P}m} kT=2.822\,kT$.
\footnote{More exactly $x_{{\rm P}m}=2.821439372$, and is a root of the
equation $3[1-\exp(-x_{{\rm P}m})]=x_{{\rm P}m}$}
It happens because the main energy is emitted when the black hole mass
was not far from the initial mass $M_0$. For black holes with
$M_0=4\cdot 10^{14}\, g$, which are evaporated to the present time, the
maximum energy of emitted gravitons (and photons), with account of (\ref{eq8}),
is about $\varepsilon_{gr,m}=x_{m{\rm b}} kT_{\rm H}\,
=0.105\,$ GeV.

The equilibrium energy spectrum $E(\varepsilon)$ of the massless fermions, or
any other ultra-relativistic fermions with a statistical weight
$g_f=2$ emitted during a black hole
evaporation (neutrino+antineutrino, electrons, positrons),
 is found analogously to (\ref{eq9})

\begin{equation}
\label{eq13}
{\cal E}(\varepsilon)=\frac{15}{79 \pi^5} \frac{\hbar c^5}{G} \frac {h}
{\varepsilon^2}\int_0^{x_0} \frac{x^4 dx}{e^x+1}\,\, {\rm erg/Hz}\,,
\end{equation}
$$x=\frac{8\pi GM_{\rm bh} \varepsilon}{\hbar c^3},\,\,\,
x_0=\frac{8\pi GM_0\, \varepsilon}{\hbar c^3}.$$
The spectrum (\ref{eq9}) has a maximum at

\begin{equation}
\label{eq14}
\varepsilon_m=\frac{\hbar c^3\, x_{m{\rm f}}}{8 \pi GM_0},
\end{equation}
where $x_{m{\rm f}}$ is a solution of the integral equation

\begin{equation}
\label{eq15}
\frac{x_{m{\rm f}}^5}{e^{x_{m{\rm f}}}+1}=2\int_0^{x_{m{\rm f}}}\frac{x^4}
{e^{x}+1}dx,\,\,\,x_{m{\rm f}}=4.350853675,\,\,\,
 \varepsilon_m=4.351\,kT_{{\rm H}0},
\end{equation}
while the maximum of the Fermi spectrum of ultrarelativistic particles with
zero chemical potential corresponds to
$\varepsilon_{{\rm F}m}=x_{{\rm F}m} kT=3.13101972\, kT$, and
$x_{{\rm F}m}$ is found from the equation
$3(1+e^{-x_{{\rm F}m}})=x_{{\rm F}m}$.

\section{Estimations of the flux of high-energy \\ gravitons}

The restrictions to the density of the primordial black holes, following from
the observations of the gamma ray background (Barrau et al., 2003),
are $\Omega_{PBH} < 3.3\cdot 10^{-9}$, $\Omega_{PBH}=\rho_{PBH}/\rho_{cr}$.
Taking (Naselskij et al, 2003) the Hubble constant $H=71$\,km/s/Mpc,
and $\rho_{cr}=1.8\cdot 10^{-29}$g/cm$^3$,
we obtain the restriction to the energy density of the gravitons emitted
by PBH as
\begin{equation}
\label{eq16}
{\cal E}_{gr}\approx \frac{8}{79}\rho_{PBH}c^2=2.7\cdot
10^{-18}{\rm ergs/cm}^3=1.7\cdot 10^{-15} {\rm GeV/cm}^{3}.
\end{equation}
The energy of the gravitons emitted during evaporation of PBH is about 0.1 GeV,
so the upper limit to the flux of the 0.1 GeV gravitons
is $F_{gr}<c{\cal E}_{gr}/4
\approx 1.3 \cdot 10^{-4}$ gravitons/cm$^2$/s.

Black holes with lower initial masses evaporate earlier and emit more
energetic particles, but due to the red shift $z$, the observed energy of
the evaporated particles at the present time
decreases with the initial mass. According to (\ref{eq1}),(\ref{eq7}),
(\ref{eq12}) the evaporation time $t_{\rm bh}$ and co-moving energy of
the maximum of the
spectrum $E_m=h\nu_m$ for gravitons and photons are

\begin{equation}
\label{eq17}
t_{\rm bh}=5\cdot 10^{17}\left(\frac{M}{4\cdot 10^{14}}\right)^3 \,{\rm s},
\quad E_m=105 \left(\frac{4\cdot 10^{14}}{M}\right)\, {\rm MeV}.
\end{equation}
Let us find the energy of the photon or
the graviton, produced at $t=t_{\rm bh}$ by black holes
evaporated at that time,
with which it comes to the present time.
We take into account, that in the matter dominated universe
at zero pressure (after recombination) we have (Peebles, 1971)

\begin{equation}
\label{eq18}
P=0, \,\,\, 1+z=\left(\frac{5\cdot
10^{17}}{t}\right)^{2/3}, \,\,\,
E_{m0}=\frac{E_m}{1+z}=105\left(\frac{M}{4\cdot 10^{14}}\right)\,\,{\rm MeV}.
\end{equation}
Before recombination in the radiation dominated universe the
following relations are valid at $z>z_{rec}$:

\begin{equation}
\label{eq19}
P=\frac{\varepsilon}{3}, \,\,\,
\frac{1+z}{1+z_{rec}}=\left(\frac{t_{rec}}{t}\right)^{1/2}, \,\,\,
E_{m,rec}=E_m\frac{1+z_{rec}}{1+z},\,\,\, \rho=\frac{4.5\cdot 10^5}{t^2}
{\rm g/cm}^3.
\end{equation}
In the universe with two stages of expansion the relations
(\ref{eq18}) and (\ref{eq19}) are valid after and before recombination.
Taking the recombination at $z_{rec}=1000$,
($t_{rec}=5\cdot 10^{17}/(1+z_{rec})^{3/2}=
1.6\cdot 10^{13}$ s, $T_{rec}=3000$ K), we obtain the ratio of the
energies at the moment of recombination and at present time
$E_{m0}=10^{-3}E_{rec}$, and the mass of a black hole evaporated at the
moment of recombination is equal to
\begin{equation}
\label{eq19a}
M_{bh,rec}=\frac{4\cdot 10^{14}}{10^{1.5}}=1.3\cdot 10^{13}\,\, {\rm g}.
\end{equation}
The formula (\ref{eq18}) for the observed energy of evaporated photons
(gravitons), in the case of free propagation, is valid in the mass interval
$1.3\cdot 10^{13}< M_{bh}<4\cdot 10^{14}$. Combining (\ref{eq17}) -
(\ref{eq19a}), we obtain the observed energy of freely propagating quanta
(gravitons) emitted by black holes evaporated to the time $t<t_{rec}$ as

\begin{equation}
\label{eq19b}
E_{m0}=\frac{105}{(1+z_{rec})^{0.25}}\left(\frac{M}{4\cdot 10^{14}}\right)^{0.5}
\,\, {\rm MeV},
\quad {\rm for} \quad M_{bh}<\frac{4\cdot 10^{14}}{10^{1.5}}\,\,{\rm g}.
\end{equation}
It was shown by Zeldovich and Sunyaev (1969) that
any extra heat introduced into the universe will be
thermalized, provided it is introduced at $z >z_{ph}= 10^6$.
The point $z=z_{ph}= 10^6$
corresponds to $t = t_{ph}= 1.6 \cdot 10^7$ s, in the universe
with two stages.
The corresponding  temperature of the universe is connected only
with the redshift during adiabatic expansion of a pure photon gas, and
is equal to $T_{ph}= 3\cdot 10^7$ K.

Black holes
evaporated at $z>z_{ph}$ do not disturb the Planckian cosmic microwave
background (CMB), but produce energetic nonthermal gravitons and neutrino.
The heat produced at $z>z_{\nu}$ is so large that neutrino thermalization
have time to occur, produce only
energetic gravitons. The values $z_{\nu}$, $t_{\nu}$, $T_{\nu}$,
may be estimated from the condition that the
mean free path of neutrino due to interaction with pairs is much smaller
than the horizon.
The number density of the electron-positron and other pairs may be estimated,
using
(\ref{eq19}), as

\begin{equation}
\label{eq20}
N_{e^+e^-}=\frac{\rho\, c^2}{x_{Fm}kT}=\frac{(\rho\, c^2)^{3/4}
a^{1/4}}{k\, x_{Fm}n^{1/4}}=\frac{4.4 \cdot 10^{31}}{t^{3/2}}.
\end{equation}
In the radiation dominant plasma $T=(\rho c^2/n\,a)^{1/4}$,
$a=7.565\cdot 10^{-15}$ is a constant of the radiation density, $n$ is
the number of sorts of low-mass and massless particles taken as $n=6$.

Estimating the cross section of neutrino interaction with electrons as
$\sigma_{\nu e}
\approx 10^{-44}$ cm$^2$, we obtain from the relation
$\frac{1}{\sigma_{\nu e} N{e^+e^-}}\sim \frac{ct}{10}
\ll ct $ the corresponding values as

\begin{equation}
\label{eq21}
t_{\nu}=2 \cdot 10^{-6}\,\,{\rm s}, \quad
z_{\nu}=3 \cdot 10^{12}, \quad
T_{\nu}=8 \cdot 10^{12}\,\, {\rm K}.
\end{equation}
The primordial black holes with masses less than $M_{ph}=
1.3\cdot 10^{11}$ g, or $M_{\nu}=6\cdot 10^6$ g,
are evaporating without disturbing CMB spectrum, or without disturbing also the
neutrino background spectrum. The  spectra maximum of gravitons emitted at
these
two moments, according to (\ref{eq1}), (\ref{eq12}), are equal
to $E_{m,ph}=320$ GeV, $E_{m,\nu}=7 \cdot 10^6$ GeV.
Using corresponding redshifts,
we obtain the following energies of the gravitons at present time:
$E_{m0,ph}=320$ keV, $E_{m0,\nu}=2$ keV.

Supposing, that considerable ($\sim 0.5$) part
of the CMB was produced by primordial black holes
evaporation, we obtain the upper limit of the flux of gravitons, produced at
different
redshifts. The CMB radiation corresponds to $\Omega \sim 7 \cdot 10^{-5}$,
which exceeds by
$2\cdot 10^4$ times the upper limit $\Omega_{PBH}$ for black holes evaporated
by the present time. The upper limits for graviton fluxes, produced by black
holes evaporated at earlier times, are equal to
\footnote{It was noted by Novikov et al. (1979), that BH with masses
$10^{10} -- 10^{13}$ g evaporated at the epoch of nucleosynthesis
influence the helium production,
so the upper limit of the graviton flux in the first line of (\ref{eq22})
decreases about an order of magnitude.}
\begin{eqnarray}
F^{(ph)}_{gr}<\sim 400 \,\,\, {\rm gravitons/cm}^2/{\rm s}\,\,\,
{\rm at}\,\,\, E^{(2)}_{m0,ph}=320\, {\rm keV},\nonumber\\
\label{eq22}
F^{(\nu)}_{gr}<\sim 6 \cdot 10^{4} \,\,\, {\rm gravitons/cm}^2/{\rm s}\,\,\,
{\rm at}\,\,\, E^{(2)}_{m0,\nu}=2\, {\rm keV}.
\end{eqnarray}
In the first case similar fluxes of neutrinos with approximately the
same energies
as gravitons are expected, in the last case neutrinos should be
thermalized. According to (\ref{eq19b}), the gravitons with energies of
optical spectrum ($\sim 1$ eV)
may be produced by the black holes with mass $\sim 1$ gramm, and since the
upper limit
for the total flux of gravitons remains the same, we have

\begin{equation}
\label{eq21a}
F^{(opt)}_{gr}<\sim 1.2 \cdot 10^{8} {\rm gravitons/cm}^2/{\rm s}.
\end{equation}
On the other hand, the optical
photons are produced by presently evaporating black holes with mass about
$5\cdot 10^{-12}M_{\odot}$,
corresponding to the Hawking temperature about 1 eV. The energy flux from
the evaporation of such black holes with the total mass equal to solar mass
and solar surface temperature is

\begin{equation}
\label{eq21b}
F^{(opt)}_{bh}=F_{\odot}\left(\frac{R_{bh}}{R_{\odot}}\right)^2\frac{M_{\odot}}
{M_{bh}}=10^{-22}F_{\odot},
\end{equation}
taking into account the relation for the gravitational radius
$R_{bh}\equiv r_g=\frac{2GM}{c^2}=3\cdot 10^5\, M/M_{\odot}$ cm.
While we cannot increase the total mass of black holes more than $\sim 50$
times compared to
the present total mass of stars, the optical flux of presently evaporated
black holes does not exceed $10^{-20}$ of total stellar optical flux, i.e.
completely negligible.
Roughly speaking, the optical stellar flux is of the order of the flux of
the microwave background, so the flux of optical gravitons from presently
evaporating black holes does not exceed $10^{-20}$ of its upper limit
in (\ref{eq21a}).

The above estimations have been done for pure equilibrium (Planck and Fermi)
spectra of evaporating particles. Calculations of Page (1976), where deviation
from equilibrium spectra had been taken into account, gave for non-rotating
black holes redistribution of the emitting particles in favor of particles
with lower spins.
The amount of radiated gravitons is about 50 times less than the sum of
photons and pairs, which are transformed into photons during thermalization.
In the equilibrium the graviton flux is about 3 times less than that  of the
sum of radiation and pairs.
Therefore, the estimations
of the graviton fluxes in (\ref{eq22}) should be reduced by a factor of 20.
In the case of
evaporation of a rapidly rotating black hole the situation is opposite,
and radiation of particles with larger spins (gravitons)
(Novikov and Frolov, 1998) prevails. In that case we may expect much lower
reducing (if any) of the graviton flux estimations in (\ref{eq22}).

\section{Perspectives of the HFGW detection }

Detection of very high frequency gravitational
radiation of cosmic origin previously did not attract a serious
attention and corresponding devices were not developed. One can
appeal only to ideas of conceivable lab detectors for the
Hertz GW experiment in the ultra high frequency radio range
$(10^{9}-10^{12})$ Hz and at optical frequencies $(10^{13}-
10^{15})$ Hz. There is nothing with respect to the GW detection
in the region of "gamma ray" frequencies.
However a rough estimate of the "detectable signal" may be given
from general principles without a consideration of a concrete
technical construction of devices.
\par
A relative energy variation of some HFGW detector $\Delta{W}/W$ has
to be proportional to metric perturbations $h$ accumulated during
the measurement time $\tau_{m}$, i.e the "signal response" is expected
as $ \Delta{W}/W \sim h\,\omega\tau_{m}$. For to be registered it
must overcome the quantum fluctuation standard
$(1/\sqrt{N})\simeq \sqrt {\hbar\omega/W}$, where $N$ is a number of
quanta of the electromagnetic pump field of the detector . From this
condition one can define a minimal registered $h_{min}$ and then a
limit for the detectable GW energy flux.

Such an approach is valid for the resonance deterministic signals
and detectors having not deal with the condition of "wave synchronism",
for example it might be electromagnetic or optical resonators of a large
volume (Grishchuk, Sazhin 1975). In the case of receiving
stochastic signals a difference consists in a more slow rate of
accumulation, i.e the accumulation factor is $\sim\sqrt{\omega \tau_{m}}$.
For a single act of measurement (one detector, one switch on-off)
the value of accumulation factor is limited by the detector relaxation
time $\sqrt{\omega \tau_{r}} \simeq\sqrt{Q}$, where $Q$ is the oscillation
quality factor. However if there is a possibility to use simultaneously
two coherent (synchronized) detectors to separate (to filter)
correlated "GW-noise signals" from independent intrinsic noises of detection
devices then the accumulation might be continued and the effective response
would grow with time as $\sqrt{\tau_{m}/\tau_{r}}$ (see for example
(Allen B 1997)) under an assumption of the permanent existence of the measured
GW noise background. Finally, the potential detection sensitivity may
be written as
\begin{equation}
h_{min}\sim \sqrt{\frac{\hbar\omega}{W}}\frac{1}{\sqrt{Q}}
\sqrt{\frac{\tau_r}{\tau_m}} \simeq
\sqrt{\frac{\hbar}{W\tau_{m}}}
\label{eq23}
\end{equation}
Thus, the principal level of the potential sensitivity is defined only by a
total energy stored in a detector and a value of allowed measurement time. \\
For numerical estimate let's take parameters used in the paper
of Grishchuk (2003) for the electromagnetic detector at frequencies
$10^{8}-10^{9}$ Hz. The electromagnetic field $E=10^{4}$ V/cm, the detector
cell volume $\lambda^{3}\simeq 10^{6}$ cm$^{3}$, the number of independent
cells $N=1000$, (then the total energy stored $W\simeq (4/25)\,10^{21}$ erg);
the measured time $ \tau_{m}=10^{7}$ sec (three months).

With these parameters the formula (\ref{eq23}) results in the estimate
$h_{min}\sim 10^{-28}$. The only conceivable way for further enhancing the
sensitivity may be associated with using of a special quantum squeezed
states of electromagnetic field inside the detector with a suppressed
variance of photon noise (Grishchuk 2003). However it is unlikely to wait
an improvement more a one-two orders of the magnitude i.e. the extremely
optimistic estimate of the detection sensitivity can not be prolonged
below the $h_{min}=10^{-30}$. Unfortunately even this level of
"delectability" is insufficient to register very high frequency
stochastic GW background. \\
In reality the average amplitude of metric perturbation integrated in the
bandwidth $\Delta \omega \sim \omega=E_{m}/\hbar$ is defined
by the spectral density of GW flux $F_{\omega}$ according to the
following relation (Landau, Lifshitz 1988)
\begin{equation}
<h>|_{\omega}\simeq\sqrt{\frac{16\pi G}{c^{3}\omega^{2}}\,
F_{\omega}\Delta \omega}
\simeq 3\,10^{-19}\omega^{-1}\,
(\frac{F_{0}}{1\,{\rm erg/sm}^{2}{\rm sec}})^{1/2}
\label{eq24}
\end{equation}
where we define $F_{0}=F_{\omega}\Delta\omega$.

For primordial gravitons produced by mini black hole evaporation
$F_{0}\sim 10^{-4}$ erg/cm$^2$ sec in the very wide frequency range,
(\ref{eq22}),(\ref{eq21a}). Then we have from (\ref{eq24}) the estimation
$<h>|_{\omega}\sim 3\,10^{-31}$ in the  "radio high frequency range"
$\omega=10^{10}$ rad/sec and much less $<h>|_{\omega}\sim 3\,10^{-36}$
at the "optical frequencies" $\omega=10^{15}$ rad/sec, i.e.$E_{h\nu}
\sim (1 - 10)$ eV.
Thus one has to conclude that a direct detection of the "black hole
gravitons" is impossible. The same is valid for the high frequency tail
of the relict GW background and thermal GW noise of stars.

A high frequency GW source has to provide the flux at least
$F_{0}\sim (1 - 10)$ erg/cm$^{2}$ sec at the Earth surface to
be detected in principle. A weak hope to get something like this
may be associated with GW grasers if they would have an extremely
sharp beaming and would be discovered nearly the Earth.
Then for their detection one could apply detectors based on the wave
synchronism of GW and optical waves in nonlinear opto-acoustical media
(Pisarev 1979, Bessonov 1998, Rudenko 2003) which looks more effective
but at present is not yet completely developed.
\par
Among the known detectors, the optical GW-interferometers with
FP cavity in the arms in principle
are able to detect the short wave GW-radiation $\lambda_g \ll L$
under a special condition of the so called "grav-optical resonance"
$(2L/\lambda_g)=k$, where $k$ is a large integer number (for example:
$k=10^3 ,L=10^5$ cm, $\lambda_g =10^{2}$ cm) (Rudenko, Sazhin 1980).
This condition means that photons meet the same gravitational
wave phase at moments of their reflection from interferometer arm
mirrors. It results in the accumulation of optical response
defined by the "finess factor" (number of photon round trips
during of the optical relaxation time) i.e. it is a similar
accumulation effect which characterizes an operation of GW
interferometers at low frequencies. So the expected sensitivity
can not be better than the advanced sensitivity of the VIRGO and LIGO
interferometers or LCGT project (Kuroda et.al.)
$10^{-22}-10^{-24}$ Hz$^{-1/2}$.

\section{Conclusions}

Analysis of high frequency GW produced by evaporation of black holes of
very small masses $1-10^7$ g have shown, that the energy density of
such GW may be of the order of the present energy density of optical
photon background produced by stars. Production of this GW does not
change any characteristics of other backgrounds, because all products of
the BH evaporation except GW should be thermalized. In the case of BH of larger
masses $10^7-10^{11}$ g we may expect, in addition to GW, the formation of
neutrino background with energies in the X-ray region $2-300$ keV for both
types of the background. Note that considerably larger energy density of
the neutrino background at energies in MeV region is produced by core-collapse
supernovae during formation of neutron stars (Bisnovatyi-Kogan and Seidov,
1982).

The present status of experimental technique does not permit us
to expect the detection
of high energy GW stochastic background in the nearest future, but it should
be beared in mind for further development. Detection of the such GW, whose
origin is
definitely could be connected with primordial BH evaporation, would give an
answer to fundamental questions of the physics of BH evaporation, the origin
of cosmic microwave background, and physical processes in the early universe.

Authors would like gratitude colleagues from MSU prof. Lipunov V.M. and
prof. Postnov K.A. for useful discussions; prof Galtsov D.V. for comments
concerning the generation of HFGW by the relativistic plasma.\\
The paper was partly supported from RFBR grant 02-02-16900 and INTAS
grant 00491.
\bigskip



\end{document}